\documentclass{iopart}
\usepackage{sublabel,epsfig,amssymb}

\renewcommand{\vec}[1]{{\bf{#1}}}

\newcommand{\citep}{\cite}

\begin{document}

\title[MOG cosmology without dark matter or $\Lambda$]{Modified Gravity: Cosmology without dark matter or Einstein's cosmological constant}

\author{J. W. Moffat$^{1,2}$ and V. T. Toth$^1$\\
$^1$Perimeter Institute for Theoretical Physics, Waterloo, Ontario N2L 2Y5, Canada\\
$^2$Department of Physics, University of Waterloo, Waterloo, Ontario N2L 3G1, Canada}

\begin{abstract}
We explore the cosmological consequences of Modified Gravity (MOG), and find that it provides, using a minimal number of parameters, good fits to data, including CMB temperature anisotropy, galaxy power spectrum, and supernova luminosity-distance observations without exotic dark matter. MOG predicts a bouncing cosmology with a vacuum energy term that yields accelerating expansion and an age of $\sim$13 billion years.
\end{abstract}

%\begin{keywords}
%{cosmology:theory -- large-scale structure of Universe -- gravitation.}
%\end{keywords}

\maketitle

\section{Introduction}

The preferred model of cosmology today, the $\Lambda$CDM model, provides an excellent fit to cosmological observations, but at a substantial cost: according to this model, {\em about 95\% of the universe is either invisible or undetectable, or possibly both} \citep{Komatsu2008}. This fact provides a strong incentive to seek alternative explanations that can account for cosmological observations without resorting to dark matter or Einstein's cosmological constant.

For gravitational theories designed to challenge the $\Lambda$CDM model, the bar is set increasingly higher by recent discoveries. Not only do such theories have to explain successfully the velocity dispersions, rotational curves, and gravitational lensing of galaxies and galaxy clusters, the theories must also be in accord with cosmological observations, notably the acoustic power spectrum of the cosmic microwave background (CMB), the matter power spectrum of galaxies, and the recent observation of the luminosity-distance relationship of high-$z$ supernovae, which is seen as evidence for ``dark energy''.

Modified Gravity (MOG) \citep{Moffat2006a} has been used successfully to account for galaxy cluster masses \citep{Brownstein2006b}, the rotation curves of galaxies \citep{Brownstein2006a}, velocity dispersions of satellite galaxies \citep{Moffat2007}, and globular clusters \citep{Moffat2007a}. It was also used to offer an explanation for the Bullet Cluster \citep{Brownstein2007} without resorting to cold dark matter.

Remarkably, MOG also meets the challenge posed by cosmological observations. In this paper, it is demonstrated that MOG produces an acoustic power spectrum, a matter power spectrum, and a luminosity-distance relationship that are in good agreement with observations, and require no dark matter nor Einstein's cosmological constant.

In the arguments presented here, we rely on simplified analytical calculations. We are not advocating these as substitutes for an accurate numerical analysis. However, a thorough numerical analysis requires significant time and resources; before these are committed, it is useful to be able to demonstrate if a theory is viable, and if the additional effort is warranted.

In the next section, we review the key features of MOG. This is followed by sections presenting detailed calculations for the luminosity-distance relationship of high-$z$ supernovae, the acoustic power spectrum of the CMB, and the galaxy power spectrum. A concluding section summarizes our results and maps out future steps.

\section{Modified Gravity Theory}

Modified Gravity (MOG) is a fully relativistic theory of gravitation that is derived from a relativistic action principle \citep{Moffat2006a} involving scalar, tensor, and vector fields. MOG has evolved as a result of investigations of Nonsymmetric Gravity Theory (NGT) \citep{Moffat1995}, and most recently, it has taken the form of Scalar-Tensor-Vector Gravity (STVG) \citep{Moffat2006a}. In the weak field approximation, STVG, NGT, and Metric-Skew-Tensor Gravity (MSTG) \citep{Moffat2005} produce similar results.

\subsection{Scalar-Tensor-Vector Gravity}
\label{sec:STVG}

Our modified gravity theory is based on postulating the existence of a massive vector field, $\phi_\mu$. The choice of a massive vector field is motivated by our desire to introduce a {\em repulsive} modification of the law of gravitation at short range. The vector field is coupled universally to matter. The theory, therefore, has three constants: in addition to the gravitational constant $G$, we must also consider the coupling constant $\omega$ that determines the coupling strength between the $\phi_\mu$ field and matter, and a further constant $\mu$ that arises as a result of considering a vector field of non-zero mass, and controls the coupling range. The theory promotes $G$, $\mu$, and $\omega$ to scalar fields, hence they are allowed to run, resulting in the following action \citep{Moffat2006a,Moffat2007e}:
\begin{equation}
S=S_G+S_\phi+S_S+S_M,
\end{equation}
where
%\begin{widetext}
\begin{eqnarray}
S_G&=&-\frac{1}{16\pi}\int\frac{1}{G}\left(R+2\Lambda\right)\sqrt{-g}~d^4x,\\
S_\phi&=&-\int\omega\left[\frac{1}{4}B^{\mu\nu}B_{\mu\nu}-\frac{1}{2}\mu^2\phi_\mu\phi^\mu+V_\phi(\phi)\right]\sqrt{-g}~d^4x,\\
S_S&=&-\int\frac{1}{G}\Bigg[\frac{1}{2}g^{\mu\nu}\left(\frac{\nabla_\mu G\nabla_\nu G}{G^2}+\frac{\nabla_\mu\mu\nabla_\nu\mu}{\mu^2}-\nabla_\mu\omega\nabla_\nu\omega\right)\nonumber\\
&&\qquad{}+\frac{V_G(G)}{G^2}+\frac{V_\mu(\mu)}{\mu^2}+V_\omega(\omega)\Bigg]\sqrt{-g}~d^4x,
\end{eqnarray}
where $S_M$ is the ``matter'' action, $B_{\mu\nu}=\partial_\mu\phi_\nu-\partial_\nu\phi_\mu$, while $V_\phi(\phi)$, $V_G(G)$, $V_\omega(\omega)$, and $V_\mu(\mu)$ denote the self-interaction potentials associated with the vector field and the three scalar fields. The symbol $\nabla_\mu$ is used to denote covariant differentiation with respect to the metric $g^{\mu\nu}$, while the symbols $R$, $\Lambda$, and $g$ represent the Ricci-scalar, the cosmological constant, and the determinant of the metric tensor, respectively. We define the Ricci tensor as
\begin{equation} R_{\mu\nu}=\partial_\alpha\Gamma^\alpha_{\mu\nu}-\partial_\nu\Gamma^\alpha_{\mu\alpha}+\Gamma^\alpha_{\mu\nu}\Gamma^\beta_{\alpha\beta}-\Gamma^\alpha_{\mu\beta}\Gamma^\beta_{\alpha\nu}.
\end{equation}
Our units are such that the speed of light, $c=1$; we use the metric signature $(+,-,-,-)$.

\subsection{The field equations}

Using the FLRW line element $ds^2=dt^2-a(t)^2[(1-kr^2)^{-1}dr^2+r^2d\Omega^2]$ with $d\Omega^2=d\theta^2+\sin^2{\theta}d\phi^2$, we obtain the Friedmann equations in the form \citep{Moffat2007e}:
\begin{eqnarray}
&&H^2+\frac{k}{a^2}=\frac{8\pi G\rho}{3}
-\frac{4\pi}{3}\left(\frac{\dot{G}^2}{G^2}+\frac{\dot{\mu}^2}{\mu^2}-\dot{\omega}^2-G\omega\mu^2\phi_0^2\right)\nonumber\\
&&\qquad{}+\frac{8\pi}{3}\left(
\omega GV_\phi+\frac{V_G}{G^2}+\frac{V_\mu}{\mu^2}+V_\omega
\right)
+\frac{\Lambda}{3}+H\frac{\dot{G}}{G},
\label{eq:FR1}\\
&&\frac{\ddot{a}}{a}=-\frac{4\pi G}{3}(\rho+3p)
+\frac{8\pi}{3}\left(\frac{\dot{G}^2}{G^2}+\frac{\dot{\mu}^2}{\mu^2}-\dot{\omega}^2-G\omega\mu^2\phi_0^2\right)\nonumber\\
&&\qquad{}+\frac{8\pi}{3}\left(
\omega GV_\phi+\frac{V_G}{G^2}+\frac{V_\mu}{\mu^2}+V_\omega
\right)
+\frac{\Lambda}{3}+H\frac{\dot{G}}{2G}+\frac{\ddot{G}}{2G}-\frac{\dot{G}^2}{G^2},
\label{eq:FR2}
\end{eqnarray}
where $H=\dot{a}/a$, and the dot denotes differentiation with respect to $t$, i.e., $\dot{y}=dy/dt$. These equations are supplemented by an additional set of three equations for the scalar fields $G$, $\mu$, and $\omega$:
\begin{eqnarray}
&&\ddot{G}+3H\dot{G}-\frac{3}{2}\frac{\dot{G}^2}{G}+\frac{G}{2}\left(\frac{\dot{\mu}^2}{\mu^2}-\dot{\omega}^2\right)+\frac{3}{G}V_G-V_G'\nonumber\\
&&\qquad{}+G\left[\frac{V_\mu}{\mu^2}+V_\omega\right]
+\frac{G}{8\pi}\Lambda-\frac{3G}{8\pi}\left(\frac{\ddot{a}}{a}+H^2\right)=0,\\
&&\ddot{\mu}+3H\dot{\mu}-\frac{\dot{\mu}^2}{\mu}-\frac{\dot{G}}{G}\dot{\mu}+G\omega\mu^3\phi_0^2+\frac{2}{\mu}V_\mu-V'_\mu=0,\\
&&\ddot{\omega}+3H\dot{\omega}-\frac{\dot{G}}{G}\dot{\omega}-\frac{1}{2}G\mu^2\phi_0^2+GV_\phi+V'_\omega=0.\label{eq:omega}
\end{eqnarray}

On the right-hand side of the Friedmann equations (\ref{eq:FR1}) and (\ref{eq:FR2}), in addition to terms describing ordinary matter and energy (characterized by $\rho$ and $p$; herein, we are only considering the case of pressureless dust, such that $p=0$ and $w=p/\rho=0$), we see several additional terms, each with its distinct equation of state. We can rewrite the Friedmann equations as
\begin{eqnarray}
&&H^2+\frac{k}{a^2}=\frac{8\pi G}{3}\left[\rho+\rho_k+\rho_V+\rho_\Lambda+\rho_G\right],\\
&&\frac{\ddot{a}}{a}=-\frac{4\pi G}{3}\bigg[\rho(1+3w)+\rho_k(1+3w_k)+\rho_V(1+3w_V)\nonumber\\
&&\qquad{}+\rho_\Lambda(1+3w_\Lambda)+\rho_G(1+3w_G)\bigg]+\frac{\ddot{G}}{2G}-\frac{\dot{G}^2}{G^2},
\end{eqnarray}
where
\begin{eqnarray}
\rho_k&=&-\frac{1}{2G}\left(\frac{\dot{G}^2}{G^2}+\frac{\dot{\mu}^2}{\mu^2}-\dot{\omega}^2-G\omega\mu^2\phi_0^2\right),\\
\rho_V&=&\frac{1}{G}\left(\omega GV_\phi+\frac{V_G}{G^2}+\frac{V_\mu}{\mu^2}+V_\omega\right),\\
\rho_\Lambda&=&\frac{\Lambda}{8\pi G},\\
\rho_G&=&\frac{3H\dot{G}}{8\pi G^2}.
\end{eqnarray}
The equations of state associated with the kinetic term $\rho_k$, potential term $\rho_V$, cosmological term $\rho_\Lambda$, and Brans-Dicke term $\rho_G$ are, respectively,
\begin{eqnarray}
w_k&=&1,\\
w_V&=&-1,\\
w_\Lambda&=&-1,\\
w_G&=&-\frac{2}{3}.
\end{eqnarray}
In particular, we note that the term associated with the potentials $V_\phi$, $V_G$, $V_\mu$, and $V_\omega$ has the equation of state $w_V=-1$, indicating that these potentials can play the role of dark energy, even if we set Einstein's cosmological constant, $\Lambda=0$.

Equations (\ref{eq:FR1}--\ref{eq:omega}) can be solved numerically, if suitable initial conditions are first established. We choose to use the following initial values:
\begin{eqnarray}
t_0&=&13.7\times 10^9~\mathrm{years},\\
a_0&=&ct_0,\\
G_0&=&6G_N,\\
\mu_0&=&a_0^{-1},\\
\omega_0&=&1/\sqrt{12},\\
\phi_0&=&0,\\
\dot{a}_0&=&H_0a_0=72a_0~\mathrm{km}/\mathrm{Mpc}~\mathrm{s},\\
\dot{G}_0=\dot{\mu}_0=\dot{\omega}_0&=&0,\\
V_G&=&0.07659537G_0^2/t_0^2,\\
V_\mu=V_\omega=V_\phi&=&0,\\
\Lambda&=&0,\\
k&=&0,
\end{eqnarray}
where $G_N$ is Newton's gravitational constant. Most of these choices are self-explanatory, and consistent with standard cosmology. Exceptions are the scalar field values. Our choices are motivated by the solution of the MOG field equations in the spherically symmetric case (for details, see \cite{Moffat2007e}), yielding a Yukawa-like modification of gravity with $\mu$ as the range parameter. We set $\mu$ to the inverse of the scale of the universe.

The value of $G_0$ is chosen to yield a model that requires no substantial amounts of nonbaryonic matter. With $G$ at 6 times the Newtonian constant, baryonic matter alone amounts to $\sim 30$\% of the critical density of the universe, hence no nonbaryonic dark matter is needed to explain a deficit in matter density.
The value of $6G_N$ is also remarkable for another reason: taking the point particle solution from the next section, an effective gravitational constant of $G_\mathrm{eff}\simeq 6G_N$ at the Yukawa distance $r=\mu^{-1}$ yields $\alpha\simeq 19$ and an effective gravitational constant of $G_\mathrm{eff}\simeq 20G_N$ at infinity. That is, on superhorizon scales our solution is consistent with an Einstein-de Sitter cosmology with no dark matter or dark energy. (This would imply a vanishing $V_G$ on superhorizon scales.)

\begin{figure}
\begin{flushright}\includegraphics[width=0.8\linewidth]{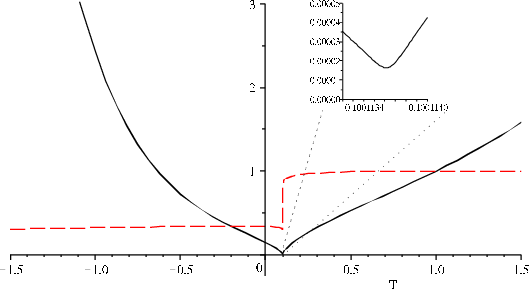}\end{flushright}
\caption{The MOG ``bouncing'' cosmology. The horizontal axis represents time, measured in Hubble units of $H_0^{-1}$. The solid (black) line is $a/a_0$, the scale factor normalized to the present epoch. The dashed (green) line is $G/G_0$. The inset shows details of the bounce, demonstrating that a smooth bounce occurs even as the matter density of the universe is more than $10^{14}$ times its present value.}
\label{fig:aG}
\end{figure}

With these initial values, the exact numerical solution of (\ref{eq:FR1}--\ref{eq:omega}) yields a bouncing cosmology (Figure~\ref{fig:aG}): the universe contracts up until approximately 0.9 times the Hubble age before the present epoch, at which time the contraction halts and expansion begins. At this time, the value of $G$ changes sharply, albeit smoothly; meanwhile, the scalar fields $\mu$ and $\omega$ remain constant.

This type of bouncing cosmology is known from the literature: Brans-Dickey theory \citep{BD1962} produces a smooth bounce if its parameter $\omega_\mathrm{BD}<-6$ \citep{GFR1973}. In our case, as the $\mu$ and $\omega$ fields are constant, our cosmology is similar to that of Brans-Dicke theory with $\omega_\mathrm{BD}=-8\pi$.

One feature not normally present in Brans-Dicke theory is the potential field $V_G$. Choosing an appropriate constant value for this potential allows us to fine-tune our cosmology, making the bounce sharper, yet keeping it smooth even as the universe achieves densities in excess of $10^{14}$ times the present density.

Furthermore, introducing a mixed equation of state for ordinary matter and radiation does not qualitatively alter this result: the essential features of the bounce as well as the time scales remain the same.

\begin{figure}
\begin{flushright}\includegraphics[width=0.8\linewidth]{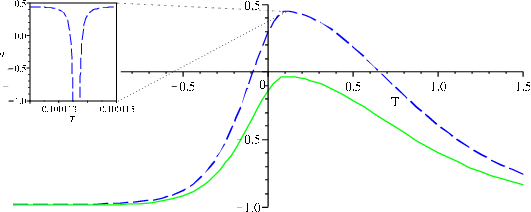}\end{flushright}
\caption{The effective equation of state $w_\mathrm{eff}$ (dotted green line) and the deceleration parameter $q$ (dashed blue line) in MOG cosmology. Inset shows the behavior of the deceleration parameter in the vicinity of the bounce. Also shown with a solid red line is the effective density normalized to the cosmic scale parameter, computed as $(G_0 a(t)^3\rho_\mathrm{eff})/(G_N a_0^3\rho_0)$.}
\label{fig:qw}
\end{figure}

To investigate further the features of the MOG cosmology, we define two additional parameters: the effective equation of state $w_\mathrm{eff}$ and the deceleration parameter $q$.

The effective equation of state is obtained by combining all terms on the right-hand side of the Friedmann equations (\ref{eq:FR1}) and (\ref{eq:FR2}) into an effective density $\rho_\mathrm{eff}$ and effective pressure $p_\mathrm{eff}=w_\mathrm{eff}\rho_\mathrm{eff}$, such that the Friedmann equations now read:
\begin{eqnarray}
H^2+\frac{k}{a^2}&=&\frac{8\pi G\rho_\mathrm{eff}}{3}+\frac{\Lambda}{3},\label{eq:FR1eff}\\
\qquad\frac{\ddot{a}}{a}&=&-\frac{4\pi}{3}(1+3w_\mathrm{eff})G\rho_\mathrm{eff}+\frac{\Lambda}{3}.\label{eq:FR2eff}
\end{eqnarray}

The deceleration parameter is defined in the usual way:
\begin{equation}
q=-\frac{\ddot{a}a}{\dot{a}^2}.
\end{equation}

These two dimensionless parameters are shown in Figure~\ref{fig:qw}. We note that the effective equation of state always remains confined between $-1\lesssim w_\mathrm{eff}\lesssim 0$, except at the bounce. In the absence of curvature ($k=0$) and a cosmological constant ($\Lambda=0$), the bounce can only occur if $\dot{a}=0$, but at the same time, $\ddot{a}$ must not be zero. This means that $p_\mathrm{eff}$ is non-zero even as $\rho_\mathrm{eff}$ is zero, as can be seen from (\ref{eq:FR1eff}) and (\ref{eq:FR2eff}). In other words, at the time of the bounce, as $q$ goes to negative infinity, so does $w_\mathrm{eff}$.

In the past, we have considered a bouncing cosmology in the MOG context \citep{Moffat2006b}. We considered the possibility that at the time of the bounce, the universe was in a state of minimal entropy, which would define a thermodynamical arrow of time that points in the $+t$ direction after the bounce, but in the $-t$ direction prior to the bounce.

\subsection{Point particles in a spherically symmetric field}

For a point particle moving in the spherically symmetric field of a gravitating source, a particularly simple solution for the acceleration is obtained \citep{Moffat2008a}:
\begin{equation}
\ddot{r}=-\frac{G_NM}{r^2}\left[1+\alpha-\alpha(1+\mu r)e^{-\mu r}\right],
\label{eq:MOGaccel}
\end{equation}
where $M$ is the source mass, while $\alpha$ determines the strength of the ``fifth force'' interaction, and $\mu$ controls its range. In prior work, $\alpha$ and $\mu$ were considered free parameters that were fitted to data. Our recent work \citep{Moffat2007e} allows us to determine $\alpha$ and $\mu$ as functions of the source mass $M$:
\begin{equation}
\alpha=\frac{M}{(\sqrt{M}+E)^2}\left(\frac{G_\infty}{G_N}-1\right),
\label{eq:alpha2}
\end{equation}
and
\begin{equation}
\mu=\frac{D}{\sqrt{M}}.
\end{equation}

This solution can be seen to satisfy the field equations in the spherically symmetric case either numerically or by deriving an approximate solution analytically \citep{Moffat2007e}. The numerical values for $D$ and $E$ are determined by matching the result against galaxy rotation curves \citep{Moffat2007e}:
\begin{eqnarray}
D&\simeq&6250~M_\odot^{1/2}\mathrm{kpc}^{-1},\label{eq:C2}\\
E&\simeq&25000~M_\odot^{1/2}.\label{eq:C1}
\end{eqnarray}
The value of $G_\infty\simeq 20G_N$ is set to ensure that at the horizon distance, the effective strength of gravity is about 6 times $G_N$, eliminating the need for cold dark matter in cosmological calculations, as described in the previous section.

\subsection{The MOG Poisson Equation}
\label{sec:Poisson}

The acceleration law (\ref{eq:MOGaccel}) is associated with the potential,
\begin{equation}
\Phi=-\frac{G_\infty M}{r}\left[1-\frac{\alpha}{1+\alpha}e^{-\mu r}\right]=\Phi_N+\Phi_Y,
\label{eq:pot}
\end{equation}
where
\begin{equation}
\Phi_N=-\frac{G_\infty M}{r}
\label{eq:PhiN}
\end{equation}
is the Newtonian gravitational potential with $G_\infty=(1+\alpha)G_N$ as the gravitational constant, and
\begin{equation}
\Phi_Y=\frac{\alpha}{1+\alpha}G_\infty M\frac{e^{-\mu r}}{r}
\label{eq:PhiM}
\end{equation}
is the Yukawa-potential. These potentials are associated with the corresponding Poisson and inhomogeneous Helmholtz equations, which are given by \citep{Brownstein2007}:
\begin{eqnarray}
\nabla^2\Phi_N(\vec{r})&=&4\pi G_\infty\rho(\vec{r}),\label{eq:PoissonN}\\
(\nabla^2-\mu^2)\Phi_Y(\vec{r})&=&-4\pi\frac{\alpha}{1+\alpha}G_\infty\rho(\vec{r}).\label{eq:PoissonM}
\end{eqnarray}
Full solutions to these potentials are given by
\begin{eqnarray}
\Phi_N(\vec{r})&=&-G_\infty\int{\frac{\rho(\vec{\tilde{r}})}{|\vec{r}-\vec{\tilde{r}}|}}~d^3\vec{\tilde{r}},\\
\Phi_Y(\vec{r})&=&\frac{\alpha}{1+\alpha}G_\infty\int{\frac{e^{-\mu|\vec{r}-\vec{\tilde{r}}|}\rho(\vec{\tilde{r}})}{|\vec{r}-\vec{\tilde{r}}|}}~d^3\vec{\tilde{r}}.
\end{eqnarray}
These solutions can be verified against Eqs.~(\ref{eq:PhiN}) and (\ref{eq:PhiM}) by applying the delta function point source density $\rho(\vec{r})=M\delta^3(\vec{r})$.

Combining Eq.~(\ref{eq:pot}) with Eqs.~(\ref{eq:PoissonN}) and (\ref{eq:PoissonM}) yields
\begin{eqnarray}
\label{eq:Poisson}
\nabla^2\Phi&=&4\pi G_N\rho(\vec{r})+\mu^2\Phi_Y(\vec{r})\\
&=&4\pi G_N\rho(\vec{r})+
\alpha\mu^2G_N\int{\frac{e^{-\mu|\vec{r}-\vec{\tilde{r}}|}\rho(\vec{\tilde{r}})}{|\vec{r}-\vec{\tilde{r}}|}}~d^3\vec{\tilde{r}},\nonumber
\end{eqnarray}
containing, in addition to the usual Newtonian term, a nonlocal source term on the right-hand side.

\section{MOG and high-{$z$} supernovae}

Type Ia supernovae are excellent standard candles for astronomy. The physics of these massive explosions is believed to be well understood, and it determines their peak luminosity. The distance to these events can be calculated from their redshift; knowing the distance and their absolute luminosity, their apparent luminosity can be computed.

\subsection{The luminosity-distance relationship}

The difference $\mu(z)$ between the absolute and apparent luminosity of a distant object can be calculated as a function of the redshift $z$, Hubble constant $H$, and cosmic deceleration $q=-\ddot{a}a/\dot{a}^2$ as \citep{Weinberg1972}:
\begin{equation}
\mu(z)=25-5\log_{10}{H}+5\log_{10}(cz)+1.086(1-q)z+...
\end{equation}

\begin{figure}
\begin{flushright}\includegraphics[width=0.8\linewidth]{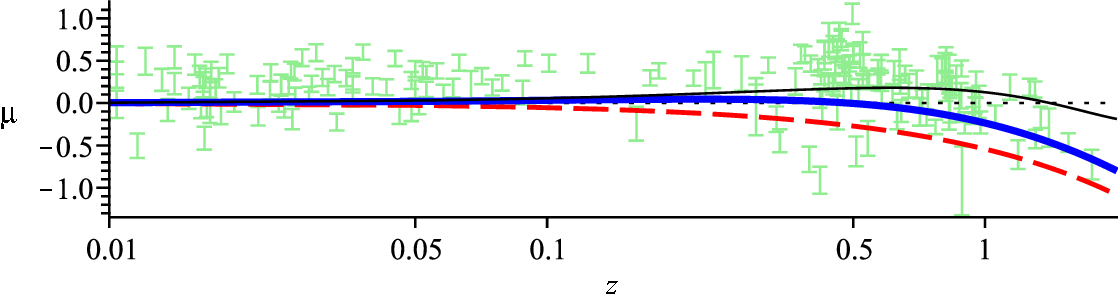}\end{flushright}
\caption{Type Ia supernova luminosity-redshift data \citep{Riess2004} and the MOG/$\Lambda$CDM predictions. No astrophysical dimming was applied. The horizontal axis corresponds to the $q=0$ empty universe. The MOG result is represented by a thick (blue) line. Dashed (red) line is a matter-dominated Einstein de-Sitter universe with $\Omega_M=1$, $q=0.5$. Thin (black) line is the $\Lambda$CDM prediction.}
\label{fig:TypeIa}
\end{figure}

Even after corrections for astrophysical dimming are applied, supernova observations are not consistent with the deceleration parameter $q=0.5$ value of a matter-dominated Einstein-de Sitter universe.

The deceleration parameter can be expressed from the Friedmann equations (\ref{eq:FR1eff}) and (\ref{eq:FR2eff}) as
\begin{equation}
q=\frac{1}{2}(1+3w_\mathrm{eff})\left(1+\frac{k}{\dot{a}^2}\right)-\frac{1}{2}(1+w_\mathrm{eff})\frac{\Lambda}{H^2},
\end{equation}
In the absence of exotic forms of matter, i.e., assuming that $w_\mathrm{eff}\ge 0$, and in the absence of curvature ($k=0$), only a positive cosmological constant, $\Lambda>0$, can reduce the value of $q$ to a value that is consistent with supernova observations.

In the case of MOG, however, the effective equation of state is $w_\mathrm{eff}<0$, and descends below $-1/3$ when the universe is about two thirds its present age. This allows $q$ to be negative, resulting in an accelerating universe even though $\Lambda=0$.

~

\subsection{Discussion}

Consistency between its predictions and supernova observations has been a major point in favor of acceptance of the $\Lambda$CDM model. However, the model leaves the question of the origin of the $\Lambda$ term open: it may be Einstein's cosmological constant, it may be a potential associated with a scalar field (quintessence), or it may be something else altogether, so long as it yields an equation of state $w<-1/3$.

MOG provides just such a term in the Friedmann equations in the form of the self-interaction potentials $V_\phi$, $V_G$, $V_\mu$ and $V_\omega$. While we have no {\it a priori} reason to single out $V_G$, the fact that it offers a means to ``fine tune'' the MOG bouncing cosmology is a major point favoring this choice. However, it must be said that interesting cosmologies can also be generated using the other self-interaction potentials. Conversely, the effects of $V_G$ in the Friedmann equations are not unique to MOG, but also present in any other theory that contains a Brans-Dicke scalar field with negative $\omega_\mathrm{BD}$ and a self-interaction potential.

At the very least, our analysis shows that MOG is consistent with supernova observations. Uncertainties in the data, and questions concerning corrections such as astrophysical dimming make it difficult to present a more definitive statement on the basis of these observations alone.

\section{MOG and the CMB}

The cosmic microwave background (CMB) is highly isotropic, showing only small temperature fluctuations as a function of sky direction. These fluctuations are not uniformly random; they show a distinct dependence on angular size, as has been demonstrated by the measurements of the Boomerang experiment \citep{Jones2006} and the Wilkinson Microwave Anisotropy Probe (WMAP) \citep{Komatsu2008}.

The angular power spectrum of the CMB can be calculated in a variety of ways. The preferred method is to use numerical software, such as {\tt CMBFAST} \citep{Seljak1996}. Unfortunately, such software packages cannot easily be adapted for use with MOG. Instead, at the present time we opt to use the excellent semi-analytical approximation developed by \cite{Mukhanov2005}. While not as accurate as numerical software, it lends itself more easily to nontrivial modifications, as the physics remains evident in the equations.

\subsection{Semi-analytical estimation of CMB anisotropies}

\cite{Mukhanov2005} calculates the correlation function $C(l)$, where $l$ is the multipole number, of the acoustic power spectrum of the CMB using the solution
\begin{equation}
\frac{C(l)}{[C(l)]_{\mathrm{low}~l}}=\frac{100}{9}(O+N),
\end{equation}
where $l\gg 1$, $O$ denotes the oscillating part of the spectrum, while the non-oscillating part is written as the sum of three parts:
\begin{equation}
N=N_1+N_2+N_3.
\end{equation}
These, in turn, are expressed as
\begin{equation}
N_1=0.063\xi^2\frac{[P-0.22(l/l_f)^{0.3}-2.6]^2}{1+0.65(l/l_f)^{1.4}}e^{-(l/l_f)^2},
\end{equation}
\begin{equation}
N_2=\frac{0.037}{(1+\xi)^{1/2}}\frac{[P-0.22(l/l_s)^{0.3}+1.7]^2}{1+0.65(l/l_s)^{1.4}}e^{-(l/l_s)^2},
\end{equation}
\begin{equation}
N_3=\frac{0.033}{(1+\xi)^{3/2}}\frac{[P-0.5(l/l_s)^{0.55}+2.2]^2}{1+2(l/l_s)^2}e^{-(l/l_s)^2}.
\end{equation}
The oscillating part of the spectrum is written as
\begin{eqnarray}
O&=&e^{-(l/l_s)^2}\sqrt{\frac{\pi}{\bar\rho l}}\nonumber\\
&\times&\left[A_1\cos{\left(\bar\rho l+\frac{\pi}{4}\right)}+A_2\cos{\left(2\bar\rho l+\frac{\pi}{4}\right)}\right],
\end{eqnarray}
where
\begin{equation}
A_1=0.1\xi\frac{(P-0.78)^2-4.3}{(1+\xi)^{1/4}}e^{\frac{1}{2}(l_s^{-2}-l_f^{-2})l^2},
\end{equation}
and
\begin{equation}
A_2=0.14\frac{(0.5+0.36P)^2}{(1+\xi)^{1/2}}.
\end{equation}
The parameters that occur in these expressions are as follows. First, the baryon density parameter:
\begin{equation}
\xi=17\left(\Omega_bh_{75}^2\right),
\label{eq:xi}
\end{equation}
where $\Omega_b\simeq 0.035$ is the baryon content of the universe at present relative to the critical density, and $h_{75}=H/(75~\mathrm{km/s/Mpc})$. The growth term of the transfer function is represented by
\begin{equation}
P=\ln{\frac{\Omega_m^{-0.09}l}{200\sqrt{\Omega_mh_{75}^2}}},
\end{equation}
where $\Omega_m\simeq 0.3$ is the total matter content (baryonic matter, neutrinos, and cold dark matter). The free-streaming and Silk damping scales are determined, respectively, by
\begin{equation}
l_f=1300\left[1+7.8\times 10^{-2}\left(\Omega_mh_{75}^2\right)^{-1}\right]^{1/2}\Omega_m^{0.09},
\label{eq:lf}
\end{equation}
\begin{equation}
l_s=\frac{0.7l_f}{\sqrt{\frac{1+0.56\xi}{1+\xi}+\frac{0.8}{\xi(1+\xi)}\frac{\left(\Omega_mh_{75}^2\right)^{1/2}}{\left[1+\left(1+\frac{100}{7.8}\Omega_mh_{75}^2\right)^{-1/2}\right]^2}}}.
\end{equation}
Lastly, the location of the acoustic peaks is determined by the parameter\footnote{Note that we slightly adjusted the coefficients of (\ref{eq:lf}) and (\ref{eq:rho}), which improved the fit noticeably, while remaining fully consistent with Mukhanov's derivation.}
\begin{equation}
\bar\rho=0.015(1+0.13\xi)^{-1}(\Omega_mh_{75}^{3.1})^{0.16}.
\label{eq:rho}
\end{equation}

\subsection{The MOG CMB spectrum}

The semi-analytical approximation presented in the previous section can be adapted to the MOG case by making two important observations.

First, in all expressions involving the value of Mukhanov's $\Omega_m$ (which includes contributions from baryonic matter and cold dark matter using Newton's gravitational constant), we need to use $\Omega_M\simeq 0.3$ (which includes baryonic matter only, using the running value of the gravitational constant, $G_\mathrm{eff}\simeq 6G_N$). Second, we notice that the value of $\Omega_b$ in (\ref{eq:xi}) {\em does not depend on the effective value of the gravitational constant}, as this value is a function of the speed of sound, which depends on the (baryonic) matter density, regardless of gravitation. In other words, $\Omega_b\simeq 0.035$ is calculated using Newton's gravitational constant.

\begin{figure}
\begin{flushright}\includegraphics[width=0.8\linewidth]{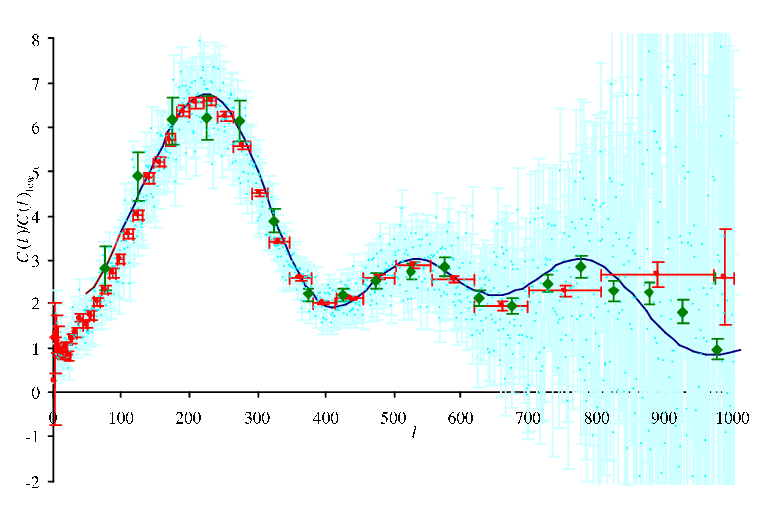}\end{flushright}
\caption{MOG and the acoustic power spectrum. Calculated using $\Omega_M=0.3$, $\Omega_b=0.035$, $H_0=71$~km/s/Mpc. Also shown are the raw WMAP 3-year data set (light blue), binned averages with horizontal and vertical error bars provided by the WMAP project (red), and data from the Boomerang experiment (green).}
\label{fig:CMB}
\end{figure}

After we modify Mukhanov's semi-analytical formulation by taking these considerations into account, we obtain the fit to the acoustic power spectrum shown in Figure~\ref{fig:CMB}.

\subsection{Discussion}

As Figure~\ref{fig:CMB} demonstrates, to the extent that Mukhanov's formulation is applicable to MOG, the theory achieves excellent agreement with the observed acoustic power spectrum. We wish to emphasize that this result was obtained without fine-tuning. The MOG constant $\mu$ was assumed to be equal to the inverse of the radius of the visible universe. Thereafter, the value of $\alpha$ is fixed if we wish to ensure $\Omega_M\simeq 0.3$. This was sufficient to achieve good agreement with the data.

\section{MOG and the matter power spectrum}

The distribution of mass in the universe is not uniform. Due to gravitational self-attraction, matter tends to ``clump'' into ever denser concentrations, leaving large voids in between. In the early universe, this process is counteracted by pressure. The process is further complicated by the fact that in the early universe, the energy density of radiation was comparable to that of matter.

\subsection{Density fluctuations in Newtonian gravity}

To first order, this process can be investigated using perturbation theory. Taking an arbitrary initial distribution, one can proceed to introduce small perturbations in the density, velocity, and acceleration fields. These lead to a second-order differential equation for the density perturbation that can be solved analytically or numerically. This yields the {\em transfer function}, which determines how an initial density distribution evolves as a function of time in the presence of small perturbations.

\subsubsection{Newtonian theory of small fluctuations}

In order to see how this theory can be developed for MOG, we must first review how the density perturbation equation is derived in the Newtonian case. Our treatment follows closely the approach presented by \cite{Weinberg1972}. We begin with three equations: the continuity equation, the Euler equation, and the Poisson equation.
\sublabon{equation}\begin{equation}
\frac{\partial\rho}{\partial t}+\nabla\cdot(\rho\vec{v})=0,
\end{equation}\begin{equation}
\frac{\partial\vec{v}}{\partial t}+(\vec{v}\cdot\nabla)\vec{v}=-\frac{1}{\rho}\nabla p+\vec{g},
\end{equation}\begin{equation}
\nabla\cdot\vec{g}=-4\pi G\rho.
\end{equation}\sublaboff{equation}

First, we perturb $\rho$, $p$, $\vec{v}$ and $\vec{g}$. Spelled out in full, we get:
\sublabon{equation}\begin{equation}
\frac{\partial(\rho+\delta\rho)}{\partial t}+\nabla\cdot[(\rho+\delta\rho)(\vec{v}+\vec{\delta v})]=0,
\end{equation}\begin{equation}
\frac{\partial(\vec{v}+\vec{\delta v})}{\partial t}+[(\vec{v}+\vec{\delta v})\cdot\nabla](\vec{v}+\vec{\delta v})
\end{equation}\[
=-\frac{1}{\rho+\delta\rho}\nabla(p+\delta p)+\vec{g}+\vec{\delta g},
\]\begin{equation}
\nabla\cdot(\vec{g}+\vec{\delta g})=-4\pi G(\rho+\delta\rho).
\end{equation}\sublaboff{equation}

Subtracting the original set of equations from the new set, using $1/(\rho+\delta\rho)=(\rho-\delta\rho)/[\rho^2-(\delta\rho)^2]=1/\rho-\delta\rho/\rho^2$, and eliminating second-order terms, we obtain
\sublabon{equation}\begin{equation}
\frac{\partial\delta\rho}{\partial t}+\nabla\cdot(\delta\rho\vec{v}+\rho\vec{\delta v})=0,
\end{equation}\begin{equation}
\frac{\partial\vec{\delta v}}{\partial t}+(\vec{v}\cdot\nabla)\vec{\delta v}+(\vec{\delta v}\cdot\nabla)\vec{v}=\frac{\delta\rho}{\rho^2}\nabla p-\frac{1}{\rho}\nabla\delta p+\vec{\delta g},
\end{equation}\begin{equation}
\nabla\cdot\vec{\delta g}=-4\pi G\delta\rho.
\end{equation}\sublaboff{equation}

A further substitution can be made by observing that $\delta p=(\delta p/\delta\rho)\delta\rho=c_s^2\delta\rho$ where $c_s^2=(\partial p/\partial\rho)_\mathrm{adiabatic}$ is the speed of sound. We can also eliminate terms by observing that the original (unperturbed) state is spatially homogeneous, hence $\nabla\rho=\nabla p=0$:
\sublabon{equation}\begin{equation}
\frac{\partial\delta\rho}{\partial t}+\vec{v}\cdot\nabla\delta\rho+\delta\rho\nabla\cdot\vec{v}+\rho\nabla\cdot\vec{\delta v}=0,
\end{equation}\begin{equation}
\frac{\partial\vec{\delta v}}{\partial t}+(\vec{v}\cdot\nabla)\vec{\delta v}+(\vec{\delta v}\cdot\nabla)\vec{v}=-\frac{c_s^2}{\rho}\nabla\delta\rho+\vec{\delta g},
\end{equation}\begin{equation}
\nabla\cdot\vec{\delta g}=-4\pi G\delta\rho.
\end{equation}\sublaboff{equation}
Now we note that $\vec{v}=H\vec{x}$, hence
\[\nabla\cdot\vec{v}=H\nabla\cdot\vec{x}=3H,\]
\[(\delta\vec{v}\cdot\nabla)\vec{v}=(\delta\vec{v}\cdot\nabla)(H\vec{x})=H(\delta\vec{v}\cdot\nabla)\vec{x}=H\delta\vec{v}.\]
Therefore,
\sublabon{equation}\begin{equation}
\frac{\partial\delta\rho}{\partial t}+\vec{v}\cdot\nabla\delta\rho+3H\delta\rho+\rho\nabla\cdot\vec{\delta v}=0,
\end{equation}\begin{equation}
\frac{\partial\vec{\delta v}}{\partial t}+(\vec{v}\cdot\nabla)\vec{\delta v}+H\vec{\delta v}=-\frac{c_s^2}{\rho}\nabla\delta\rho+\vec{\delta g},
\end{equation}\begin{equation}
\nabla\cdot\vec{\delta g}=-4\pi G\delta\rho.
\end{equation}\sublaboff{equation}
The next step is a change of spatial coordinates to coordinates comoving with the Hubble flow:
\[\vec{x}=a(t)\vec{q}.\]
This means
\[\left(\frac{\partial}{\partial t}\right)_\vec{q}=\left(\frac{\partial}{\partial t}\right)_\vec{x}+\vec{v}\nabla_\vec{x},\]
and
\[\nabla_\vec{q}=a\nabla_\vec{x}.\]
After this change of coordinates, our system of equations becomes
\sublabon{equation}\begin{equation}
\frac{\partial\delta\rho}{\partial t}+3H\delta\rho+\frac{1}{a}\rho\nabla\cdot\vec{\delta v}=0\label{eq:7a},
\end{equation}\begin{equation}
\frac{\partial\vec{\delta v}}{\partial t}+H\vec{\delta v}=-\frac{c_s^2}{a\rho}\nabla\delta\rho+\vec{\delta g}\label{eq:7b},
\end{equation}\begin{equation}
\nabla\cdot\vec{\delta g}=-4\pi aG\delta\rho.\label{eq:7c}
\end{equation}\sublaboff{equation}
Now is the time to introduce the fractional amplitude $\delta=\delta\rho/\rho$. Dividing (\ref{eq:7a}) with $\rho$, we get
\begin{equation}
\dot\delta+\frac{\dot\rho}{\rho}\delta+3H\delta+\frac{1}{a}\nabla\cdot\vec{\delta v}=0.
\end{equation}
However, since $\rho=\rho_0a_0^3/a^3$, and hence $\dot\rho/\rho=-3\dot{a}/a$, the second and third terms cancel, to give
\begin{equation}
-a\dot\delta=\nabla\vec{\delta v}.
\end{equation}
Taking the gradient of (\ref{eq:7b}) and using (\ref{eq:7c}) to express $\nabla\cdot\vec{\delta g}$, we get
\begin{equation}
\frac{\partial}{\partial t}(-a\dot\delta)+H(-a\dot\delta)=-\frac{c_s^2}{a}\nabla^2\delta-4\pi Ga\rho\delta.
\end{equation}
Spelling out the derivatives, and dividing both sides with $a$, we obtain
\begin{equation}
\ddot\delta+2H\dot\delta-\frac{c_s^2}{a^2}\nabla^2\delta-4\pi G\rho\delta=0.\label{eq:delta}
\end{equation}
For every Fourier mode $\delta=\delta_\vec{k}(t)e^{i\vec{k}\cdot\vec{q}}$ (such that $\nabla^2\delta=-k^2\delta$), this gives
\begin{equation}
\ddot\delta_\vec{k}+2H\dot\delta_\vec{k}+\left(\frac{c_s^2k^2}{a^2}-4\pi G\rho\right)\delta_\vec{k}=0.
\label{eq:ddotdelta}
\end{equation}
The quantity $k/a$ is called the co-moving wave number.

If $k$ is large, solutions to (\ref{eq:ddotdelta}) are dominated by an oscillatory term; for small $k$, a growth term predominates.

A solution to (\ref{eq:ddotdelta}) tells us how a power spectrum evolves over time, as a function of the wave number; it does not specify the initial power spectrum. For this reason, solutions to (\ref{eq:ddotdelta}) are typically written in the form of a transfer function
\begin{equation}
T(k)=\frac{\delta_k(z=0)\delta_0(z=\infty)}{\delta_k(z=\infty)\delta_0(z=0)}.
\end{equation}

If the initial power spectrum and the transfer function are known, the power spectrum at a later time can be calculated (without accounting for small effects) as
\begin{equation}
P(k)=T^2(k)P_0(k).
\end{equation}

$P(k)$ is a dimensioned quantity. It is possible to form the dimensionless power spectrum
\begin{equation}
\Delta^2(k)=Ak^3T^2(k)P_0(k),
\end{equation}
where $A$ is a normalization constant determined by observation. This form often appears in the literature. In the present work, however, we are using $P(k)$ instead of $\Delta(k)$.

The initial power spectrum is believed to be a scale invariant power spectrum:
\begin{equation}
P_0(k)\propto k^n,
\end{equation}
where $n\simeq 1.$ A recent estimate on $n$ is $n=0.963^{+0.014}_{-0.015}$ \citep{Komatsu2008}.

\subsubsection{Analytical approximation}

Eq.~(\ref{eq:ddotdelta}) is not difficult to solve in principle. The solution can be written as the sum of oscillatory and growing terms. The usual physical interpretation is that when pressure is sufficient to counteract gravitational attraction, this mechanism prevents the growth of density fluctuations, and their energy is dissipated instead in the form of sound waves. When the pressure is low, however, the growth term dominates and fluctuations grow. Put into the context of an expanding universe, one can conclude that in the early stages, when the universe was hot and dense, the oscillatory term had to dominate. Later, the growth term took over, the perturbation spectrum ``froze'', affected only by uniform growth afterwards.

In practice, several issues complicate the problem. First, the early universe cannot be modeled by matter alone; it contained a mix of matter and radiation (and, possibly, neutrinos and cold dark matter.) To correctly describe this case even using the linear perturbation theory outlined in the previous sections, one needs to resort to a system of coupled differential equations describing the different mediums. Second, if the perturbations are sufficiently strong, linear theory may no longer be valid. Third, other nonlinear effects, including Silk-damping \citep{Padmanabhan1993}, cannot be excluded as their contribution is significant (indeed, Silk damping at higher wave numbers is one of the reasons why a baryon-only cosmological model based on Einstein's theory of gravity fails to account for the matter power spectrum.)

The authors of \cite{EH1998} addressed all these issues when they developed a semi-analytical solution to the baryon transfer function. This solution reportedly yields good results in the full range of $0\leq\Omega_b\leq1$. Furthermore, unlike other approximations and numerical software codes, this approach keeps the essential physics transparent, allowing us to adapt the formulation to the MOG case.

The authors of \cite{EH1998} write the transfer function as the sum of a baryonic term $T_b$ and a cold dark matter term $T_c$:
\begin{equation}
T(k)=\frac{\Omega_b}{\Omega_m}T_b(k)+\frac{\Omega_c}{\Omega_m}T_c(k),
\end{equation}
where $\Omega_c$ represents the cold dark matter content of the universe relative to the critical density. As we are investigating a cosmology with no cold dark matter, we ignore $T_c$. The baryonic part of the transfer function departs from the cold dark matter case on scales comparable to, or smaller than, the sound horizon. Consequently, the baryonic transfer function is written as
\begin{equation}
T_b(k)=\left[\frac{\tilde{T}_0(k,1,1)}{1+(ks/5.2)^2}+\frac{\alpha_be^{-(k/k_\mathrm{Silk})^{1.4}}}{1+(\beta_b/ks)^3}\right]\frac{\sin{k\tilde{s}}}{k\tilde{s}},
\label{eq:Tb}
\end{equation}
with
\begin{equation}
\tilde{T}_0(k,\alpha_c,\beta_c)=\frac{\ln{(e+1.8\beta_c\bar{q})}}{\ln{(e+1.8\beta_c\bar{q})}+C\bar{q}^2},
\end{equation}
where
\begin{equation}
C=\frac{14.2}{\alpha_c}+\frac{386}{1+69.9\bar{q}^{1.08}},
\end{equation}
and
\begin{equation}
\bar{q}=k\Theta_{2.7}^2\left(\Omega_mh^2\right)^{-1}.
\end{equation}
The sound horizon is calculated as
\begin{equation}
s=\frac{2}{3k_\mathrm{eq}}\sqrt{\frac{G}{R_\mathrm{eq}}}\ln{\frac{\sqrt{1+R_d}+\sqrt{R_d+R_\mathrm{eq}}}{1+\sqrt{R_\mathrm{eq}}}}.
\end{equation}
The scale at the equalization epoch is calculated as
\begin{equation}
k_\mathrm{eq}=7.46\times 10^{-2}\Omega_mh^2\Theta_{2.7}^{-2}.
\end{equation}
The transition from a radiation-dominated to a matter-dominated era happens at the redshift
\begin{equation}
z_\mathrm{eq}=25000\Omega_mh^2\Theta_{2.7}^{-4},
\end{equation}
while the drag era is defined as
\begin{equation}
z_d=1291\frac{(\Omega_mh^2)^{0.251}}{1+0.659(\Omega_mh^2)^{0.828}}[1+b_1(\Omega_mh^2)^{b_2}],
\end{equation}
where
\begin{equation}
b_1=0.313(\Omega_mh^2)^{-0.419}[1+0.607(\Omega_mh^2)^{0.674}],
\end{equation}
and
\begin{equation}
b_2=0.238(\Omega_mh^2)^{0.223}.
\end{equation}
The baryon-to-photon density ratio at a given redshift is calculated as
\begin{equation}
R=31.5\Omega_mh^2\Theta_{2.7}^{-4}\frac{1000}{z}.
\end{equation}
The Silk damping scale is obtained using
\begin{equation}
k_\mathrm{Silk}=1.6(\Omega_bh^2)^{0.52}(\Omega_mh^2)^{0.73}[1+(10.4\Omega_mh^2)^{-0.95}].\label{eq:Silk}
\end{equation}
The coefficients in the second term of the baryonic transfer function are written as
\begin{equation}
\alpha_b=2.07k_\mathrm{eq}s(1+R_d)^{-3/4}F\left(\frac{1+z_\mathrm{eq}}{1+z_d}\right),
\end{equation}
\begin{equation}
\beta_b=0.5+\frac{\Omega_b}{\Omega_m}+\left(3-2\frac{\Omega_b}{\Omega_m}\right)\sqrt{(17.2\Omega_mh^2)^2+1},
\end{equation}
where we used the function
\begin{equation}
F(y)=y\left[-6\sqrt{1+y}+(2+3y)\ln{\frac{\sqrt{1+y}+1}{\sqrt{1+y}-1}}\right].
\end{equation}
A shifting of nodes in the baryonic transfer function is accounted for by the quantity
\begin{equation}
\tilde{s}(k)=\frac{s}{\left[1+\left(\beta_\mathrm{node}/ks\right)^3\right]^{1/3}},
\end{equation}
where
\begin{equation}
\beta_\mathrm{node}=8.41(\Omega_mh^2)^{0.435}.
\end{equation}

The symbol $\Theta_{2.7}=T/2.7$ is the temperature of the CMB relative to 2.7~K, while $h=H/(100~\mathrm{km/s/Mpc})$. The wave number $k$ is in units of Mpc$^{-1}$.

\subsection{Density fluctuations in Modified Gravity}

\label{sec:MOGdensity}

\begin{figure}
\begin{flushright}\includegraphics[width=0.8\linewidth]{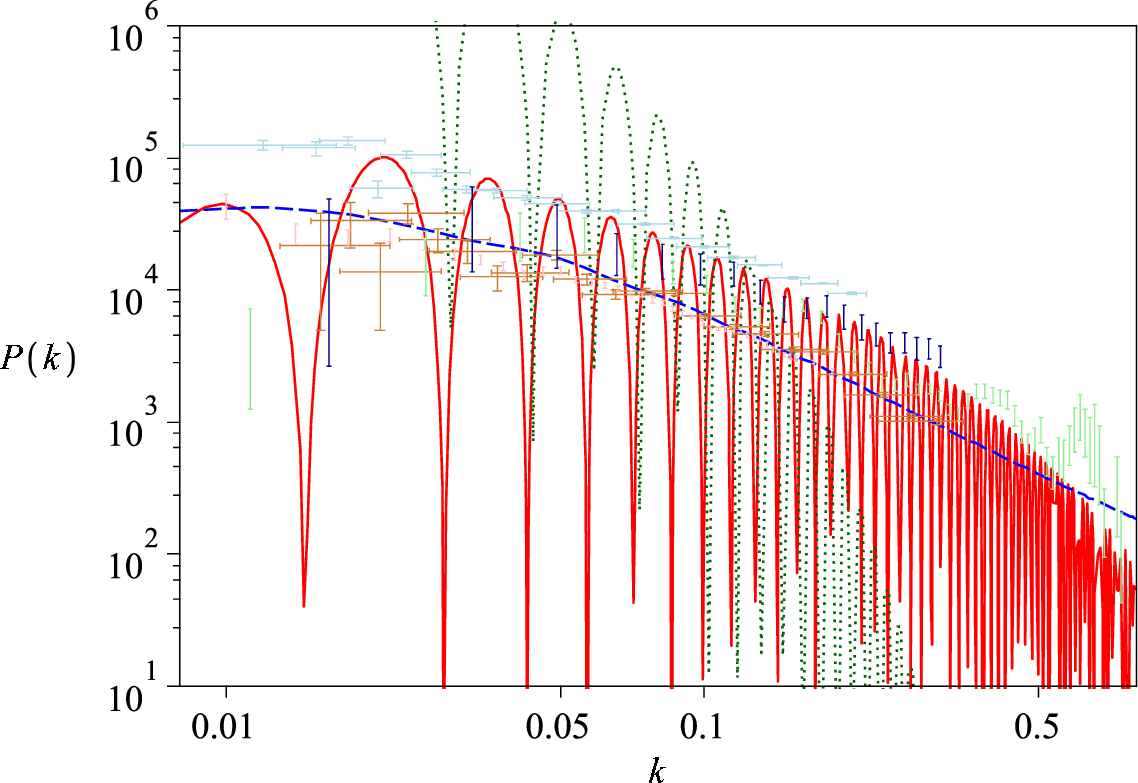}\end{flushright}
\caption{The matter power spectrum. Three models are compared against five data sets (see text): $\Lambda$CDM (dashed blue line, $\Omega_b=0.035$, $\Omega_c=0.245$, $\Omega_\Lambda=0.72$, $H=71$~km/s/Mpc), a baryon-only model (dotted green line, $\Omega_b=0.035$, $H=71$~km/s/Mpc), and MOG (solid red line, $\alpha=19$, $\mu=5h$~Mpc$^{-1}$, $\Omega_b=0.035$, $H=71$~km/s/Mpc.) Data points are colored light blue (SDSS 2006), gold (SDSS 2004), pink (2dF), light green (UKST), and dark blue (CfA).}
\label{fig:Pk}
\end{figure}

We consider the MOG Poisson equation (\ref{eq:Poisson}), established in section~\ref{sec:Poisson}. As the initial unperturbed distribution is assumed to be homogeneous, $\rho$ is not a function of $\vec{r}$ and can be taken outside the integral sign:
\begin{equation}
\Phi_Y(\vec{r})=G_N\alpha\rho\int\frac{1}{|\vec{r}-\vec{r}'|}e^{-\mu|\vec{r}-\vec{r}'|}d^3\vec{r}'.
\end{equation}
Varying $\rho$, we get
\begin{equation}
\nabla\cdot\delta\vec{g}(\vec{r})=-4\pi G_N\delta\rho(\vec{r})-\mu^2G_N\alpha\delta\rho\int\frac{1}{|\vec{r}-\vec{r}'|}e^{-\mu|\vec{r}-\vec{r}'|}d^3\vec{r}'.
\end{equation}
Accordingly, (\ref{eq:delta}) now reads
\begin{equation}
\ddot\delta+2H\dot\delta-\frac{c_s^2}{a^2}\nabla^2\delta-4\pi G_N\rho\delta-\mu^2G_N\alpha\rho\delta\int\frac{e^{-\mu|\vec{r}-\vec{r}'|}}{|\vec{r}-\vec{r}'|}d^3\vec{r}'=0.
\label{eq:19}
\end{equation}
The integral can be readily calculated. Assuming that $|\vec{r}-\vec{r}'|$ runs from 0 to the comoving wavelength $a/k$, we get
\begin{eqnarray}
\int\frac{e^{-\mu|\vec{r}-\vec{r}'|}}{|\vec{r}-\vec{r}'|}d^3\vec{r}'&=&2\int\limits_0^{\pi/2}\int\limits_0^{2\pi}\int\limits_0^{a/k}\frac{e^{-\mu r}}{r}~r^2\sin{\theta}~dr~d\phi~d\theta\nonumber\\
&=&\frac{4\pi\left[1-(1+\mu a/k)e^{-\mu a/k}\right]}{\mu^2}.
\end{eqnarray}
Substituting into (\ref{eq:19}), we get
\begin{eqnarray}
\ddot\delta+2H\dot\delta-\frac{c_s^2}{a^2}\nabla^2\delta-4\pi G_N\rho\delta&&\nonumber\\
{}-4\pi G_N\alpha\left[1-\left(1+\frac{\mu a}{k}\right)e^{-\mu a/k}\right]\rho\delta&=&0,
\end{eqnarray}
or
\begin{eqnarray}
\ddot\delta+2H\dot\delta-\frac{c_s^2}{a^2}\nabla^2\delta&&\\
-4\pi G_N\left\{1+\alpha\left[1-\left(1+\frac{\mu a}{k}\right)e^{-\mu a/k}\right]\right\}\rho\delta&=&0.\nonumber
\end{eqnarray}

This demonstrates how the effective gravitational constant
\begin{equation}
G_\mathrm{eff}=G_N\left\{1+\alpha\left[1-\left(1+\frac{\mu a}{k}\right)e^{-\mu a/k}\right]\right\}
\end{equation}
depends on the wave number.

Using $G_\mathrm{eff}$, we can express the perturbation equation as
\begin{equation}
\ddot\delta_\vec{k}+2H\dot\delta_\vec{k}+\left(\frac{c_s^2k^2}{a^2}-4\pi G_\mathrm{eff}\rho\right)\delta_\vec{k}=0.
\label{eq:MOGdelta}
\end{equation}

As the wave number $k$ appears only in the source term
$\left(\frac{c_s^2k^2}{a^2}-4\pi G_\mathrm{eff}\rho\right),$
it is easy to see that any solution of (\ref{eq:ddotdelta}) is also a solution of (\ref{eq:MOGdelta}), provided that $k$ is replaced by $k'$ in accordance with the following prescription:
\begin{equation}
k'^2=k^2+4\pi a^2\left(\frac{G_\mathrm{eff}-G_N}{G_N}\right)\lambda_J^{-2},
\end{equation}
where $\lambda_J=\sqrt{c_s^2/G_N\rho}$ is the Jeans wavelength.

This shifting of the wave number applies to the growth term of the baryonic transfer function (\ref{eq:Tb}). However, as the sound horizon scale is not affected by changes in the effective gravitational constant, terms containing $ks$ must remain unchanged. Furthermore, the Silk damping scale must also change as a result of changing gravity; this change is proportional to the 3/4th power of $G$, as demonstrated by \cite{Padmanabhan1993}, thus $k'_\mathrm{Silk}=k_\mathrm{Silk}(G/G_N)^{3/4}$ (note also Eq. \ref{eq:Silk}). Using these considerations, we obtain the modified baryonic transfer function
\begin{equation}
T'_b(k)=\frac{\sin{k\tilde{s}}}{k\tilde{s}}\times\left\{\frac{\tilde{T}_0(k',1,1)}{1+(ks/5.2)^2}+\frac{\alpha_b\exp\left({-[k/k'_\mathrm{Silk}]^{1.4}}\right)}{1+(\beta_b/ks)^3}\right\}.
\end{equation}

The effects of these changes can be summed up as follows. At low values of $k$, the transfer function is suppressed. At high values of $k$, where the transfer function is usually suppressed by Silk damping, the effect of this suppression is reduced. The combined result is that the tilt of the transfer function changes, such that its peaks are now approximately in agreement with data points, as seen in Figure~\ref{fig:Pk}.

Data points shown in this figure come from several sources. First and foremost, the two data releases of the Sloan Digital Sky Survey (SDSS) \citep{SDSS2004, SDSS2006} are presented. Additionally, data from the 2dF Galaxy Redshift Survey \citep{2dF2006}, UKST \citep{UKST1999}, and CfA130 \citep{CfA1994} surveys are shown. Apart from normalization issues, the data from these surveys are consistent in the range of $0.01~h~\mathrm{Mpc}^{-1}\leq k\leq 0.5~h~\mathrm{Mpc}^{-1}$. Some surveys provide data points outside this range, but they are not in agreement with each other.

\subsection{Discussion}

As a result of the combined effects of dampened structure growth at low values of $k$ and reduced Silk damping at high values of $k$, the slope of the MOG transfer function differs significantly from the slope of the baryonic transfer function, and matches closely the observed values of the matter power spectrum. On the other hand, the predictions of MOG and $\Lambda$CDM cosmology differ in fundamental ways.

\begin{figure}
\begin{flushright}\includegraphics[width=0.8\linewidth]{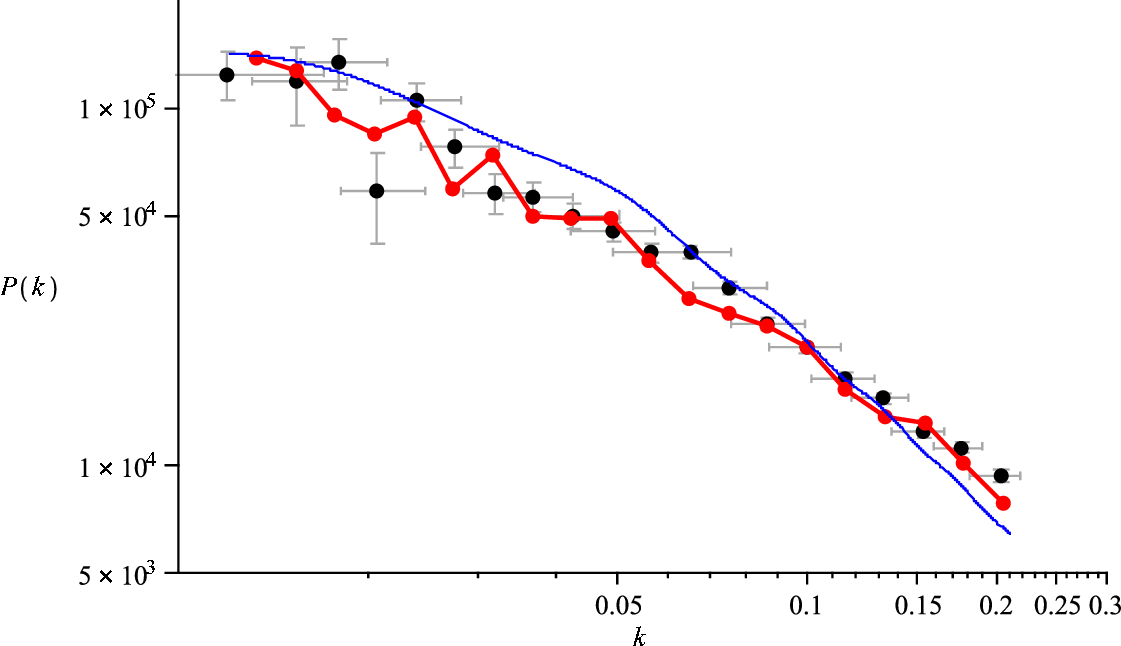}\end{flushright}
\caption{The effect of window functions on the power spectrum is demonstrated by applying the SDSS luminous red galaxy survey window functions to the MOG prediction. Baryonic oscillations are greatly dampened in the resulting curve (solid red line), yielding excellent agreement with the data after normalization. A normalized linear $\Lambda$CDM estimate is also shown (thin blue line) for comparison.}
\label{fig:baryon}
\end{figure}

First, MOG predicts oscillations in the power spectrum, which are not smoothed out by dark matter. However, the finite size of samples and the associated window functions used to produce power spectra are likely masking any such oscillations in presently available survey data (see Figure~\ref{fig:baryon}). These oscillations may be detectable in future galaxy surveys that utilize a large enough number of galaxies, and sufficiently narrow window functions in order to be sensitive to such fluctuations.

Second, MOG predicts a dampened power spectrum at both high and low values of $k$ relative to $\Lambda$CDM. Observations at sufficiently high values of $k$ may not be practical, as we are entering sub-galactic length scales. Low values of $k$ are a different matter: as accurate three-dimensional information becomes available on ever more distant galaxies, power spectrum observations are likely to be extended in this direction.

In the present work, we made no attempt to account for the possibility of a non-zero neutrino mass, and its effects on the power spectrum. Given the uncertainties in the semi-analytical approximations that we utilized, such an attempt would not have been very fruitful. Future numerical work, however, must take into account the possibility of a non-negligible contribution of neutrinos to the matter density.

~\par

\section{Conclusions}

In this paper, we demonstrated how Modified Gravity can account for key cosmological observations using a minimum number of free parameters. Although MOG permits the running of its coupling and scaling constants with time and space, we made very little use of this fact. Throughout these calculations, we used consistently the value of $\alpha\simeq 19$ for the MOG coupling constant, consistent with a flat universe with $\Omega_b\simeq 0.05$ visible matter content, no dark matter, nor Einstein's cosmological constant. In nearly all cosmological calculations, we set the MOG scaling constant $\mu$ to the inverse of the radius of the visible universe, which is a natural choice. The only exception is the mass power spectrum calculation, where the scaling constant enters in conjunction with the wave number $k$, and describes gravitational interactions between nearby concentrations of matter, not on the cosmological scale.

The theory requires {\em no other parameters} to obtain the remarkable fits to data that have been demonstrated here.

At all times, $\lim\limits_{r\rightarrow 0}G=G_N$, i.e., the effective gravitational constant at short distances remains Newton's constant of gravitation. For this reason, the predictions of MOG are {\em not contradicting our knowledge of the processes of the initial nucleosynthesis}, taking place at redshifts of $z\simeq 10^{10}$, since the interactions take place over distance scales that are much shorter than the horizon scale.

Our calculations relied on analytical approximations. This is dictated by necessity, not preference. We recognize that numerical methods, including high-accuracy solutions of coupled systems of differential equations, as in {\tt CMBFAST} \citep{Seljak1996}, or $N$-body simulations, can provide superior results, and may indeed help either to confirm or to falsify the results presented here. Nevertheless, our present work demonstrates that at the very least, MOG provides a worthy alternative to $\Lambda$CDM cosmology.

\section*{Acknowledgments}

We would like to thank John Peacock for helpful comments and suggestions.
The research was partially supported by National Research Council of Canada. Research at the Perimeter Institute for Theoretical Physics is supported by the Government of Canada through NSERC and by the Province of Ontario through the Ministry of Research and Innovation (MRI).

\bibliography{refs}
\bibliographystyle{unsrt}

\end{document}